\documentclass[floatfix,superscriptaddress,a4paper,
               showpacs,showkeys,nofootinbib,preprint]{revtex4-1}
\usepackage{epsfig}
\usepackage{latexsym}
\usepackage{xspace}
\usepackage{hyperref}
\usepackage[latin2]{inputenc}
\usepackage{indentfirst}
\usepackage{enumerate}
\usepackage{color}

\usepackage{amsmath}
\usepackage{amssymb}
\usepackage[english]{babel}
\usepackage{url}
\newcommand{\eq}[1]{\begin{align} #1 \end{align}}

\begin{document}

\title{
 Monte Carlo approach to the excluded-volume hadron resonance gas\\
 in grand canonical and canonical ensembles
}
\author{Volodymyr Vovchenko}
\affiliation{
Institut f\"ur Theoretische Physik,
Goethe Universit\"at Frankfurt, D-60438 Frankfurt am Main, Germany}
\affiliation{
Frankfurt Institute for Advanced Studies, Giersch Science Center, Campus Riedberg,
D-60438 Frankfurt am Main, Germany}
\author{Mark I. Gorenstein}
\affiliation{
Bogolyubov Institute for Theoretical Physics, 03680 Kiev, Ukraine}
\affiliation{
Frankfurt Institute for Advanced Studies, Giersch Science Center, Campus Riedberg,
D-60438 Frankfurt am Main, Germany}
\author{Horst Stoecker}
\affiliation{
Institut f\"ur Theoretische Physik,
Goethe Universit\"at Frankfurt, D-60438 Frankfurt am Main, Germany}
\affiliation{
Frankfurt Institute for Advanced Studies, Giersch Science Center, Campus Riedberg,
D-60438 Frankfurt am Main, Germany}
\affiliation{
GSI Helmholtzzentrum f\"ur Schwerionenforschung GmbH, D-64291 Darmstadt, Germany}

\date{\today}

\begin{abstract}
The Monte Carlo (MC) procedure for sampling the hadron yields within the hadron
resonance gas (HRG) model is presented. The effects of excluded-volume  due
to the finite hadron
eigenvolumes and of exact charge conservation within the canonical ensemble (CE) formulation
are simultaneously taken into account with the help of the importance sampling technique combined with the rejection sampling.
The MC procedure allows one to calculate arbitrary moments of the event-by-event hadron yields.
Note that the CE formulation for the excluded-volume HRG has not been considered before.
The MC simulations coincide
with the known analytic results in the thermodynamic limit
for the excluded-volume HRG in the grand canonical ensemble
and for the CE of non-interacting particles.
The MC procedure is applied to study the simultaneous excluded-volume
and CE effects. These effects are considered within the full HRG to calculate the particle number fluctuations and to estimate the finite size effects.
\end{abstract}

\pacs{25.75.Ag, 24.10.Pa}

\keywords{Monte Carlo, hadron resonance gas, excluded volume,  canonical ensemble}

\maketitle

\section{Introduction}
The hadron resonance gas (HRG) model denotes a class of  popular simple
models used to describe the thermodynamic properties of  QCD in the region
of temperature and baryon chemical potential where the hadronic
degrees of freedom dominate.
The HRG models give rather successful descriptions of different heavy-ion hadron yield data
over a wide range of collision energies
~\cite{CleymansSatz,CleymansRedlich1998,CleymansRedlich1999,
Becattini2001,Rafelski2003,Becattini2004,ABS2006,ABS2009,VBG2015}.
The HRG models have also been compared to and validated through the 
lattice QCD data,
both for the thermodynamical functions
of the hadron systems~\cite{Borsanyi2013,Bazavov2014}
and for fluctuations and correlations
of the conserved charges~\cite{Borsanyi:2011sw,Bazavov:2012jq}.

In its simplest form, the HRG system can be modeled as a multi-component gas
of non-interacting hadrons and resonances. One usually  refers to such a model as the ideal HRG (I-HRG) model.
It is argued~\cite{DMB} that by including
resonances
into the model, one can effectively include the interaction between the hadrons.
Short-range repulsive interactions
are usually considered within the excluded-volume~(EV)  approach.
The thermodynamically consistent procedure
to include hadron eigenvolumes
was developed in Ref.~\cite{Rischke1991}, and was often
used in fits of the HRG model of chemical freeze-out
properties~\cite{Yen1997,Gorenstein1999,PHS1999,Begun2013},
as well as for comparisons to  lattice 
QCD data~\cite{EV-latt-1,Bhattacharyya:2013oya,EV-latt-3,EV-latt-2,Vovchenko:2017xad,Alba:2017bbr}.
The importance of the excluded-volume effects in a gas of glueballs in the Yang-Mills theory was recently pointed out as well~\cite{Alba:2016fku}.
Most analyzes which employ the excluded-volume HRG
assume that all hadrons have the same eigenvolumes. However,
the eigenvolume effects essentially cancel out in the hadron yield ratios and, thus,
cannot affect  the fit quality or the extracted values of
the intensive chemical ''freeze-out parameters'', such as temperatures or chemical potentials.
Recently it has been pointed out that thermal fits are extremely sensitive to the choice of
different hadron eigenvolumes for different hadrons~\cite{Vovchenko:2015cbk,Vovchenko:2016ebv}.
If different mass-volume relations for strange and non-strange hadrons
are employed, a remarkable improvement of the fit quality of hadron yield data
can be achieved over a large range of collision energies~\cite{Alba:2016hwx}.
These eigenvolume HRG models are therefore particularly interesting.

Conserved charges are
conserved only on average in the grand canonical ensemble (GCE), but
differ from one microscopic state to another.
The exact conservation of the conserved
charges becomes important for smaller systems. Such exact  conservation of charges can
be enforced  within the canonical ensemble (CE)  \cite{CE}.
The CE formulation of the I-HRG was
successfully used to describe the hadron production data in small systems,
such as (anti)proton-proton and $e^+e^-$ collisions~\cite{Becattini1995,Becattini1997,Becattini2000,Becattini2010,VBG2015}.
The CE
strongly influences the strange \cite{CE-strange} and charm \cite{CE-charm} hadron multiplicities
as the average total numbers of strange and charm charges
are often not large (of the order of unity or smaller). It should be also noted that
for systems of non-interacting particles the CE effects
lead to noticeable suppression of particle number fluctuations for statistical systems
even in the thermodynamic limit \cite{Begun2004,Gor-2015}.

To the best of our knowledge, the CE formulation for the excluded-volume HRG is presently missing. Thus, the influence
of EV effects on the thermodynamic properties within
the CE was never explored. In the present paper a
Monte Carlo (MC) procedure is proposed which allows doing exactly that.

The paper is organized as follows. 
Different versions of the excluded-volume
models are considered in the GCE and the CE in Sections~\ref{sec-an} and \ref{sec-CE}, respectively.
In Section~\ref{sec-MC} the MC procedure is developed to calculate  the moments of particle number distributions
for the excluded-volume models, both in the GCE and CE. Section~\ref{sec-calc} presents the numerical
results, and Section~\ref{sec-sum} summarizes the paper.

\section{Excluded-Volume models in the GCE
}\label{sec-an}
Our consideration will be restricted to the case of classical
(Boltzmann) statistics. It is useful to define the single-particle
function:
\eq{\label{z}
z_i(T)~=~
\frac{g_i}{2\pi^2}\int_0^{\infty}k^2dk~
\exp\left[-~\frac{(k^2+m_i^2)^{1/2}}{T}\right]~,
}
where $g_i$ and $m_i$ are the $i$th particle degeneracy factor and mass, respectively, and $T$
is the system temperature.
In the single-component  system
the ideal gas GCE partition function reads ($z_i\equiv z$):
\eq{\label{Zid}
\mathcal{Z}_{\rm id}(V,T,\mu)=
\sum_{N=0}^{\infty} ~ \exp\left(\frac{\mu N}{T}\right)\frac{(z\,V)^N}{N!}~=~\exp(e^{\mu/T} z\,V)~,
}
where
$V$ is the total volume of the system and $\mu$ is the chemical potential.
The number of particles $N$ is fixed in the CE, and
has the Poisson distribution, $P(N)=\langle N\rangle^N \exp(-\langle N\rangle)/N!$,
in the GCE with average value $\langle N\rangle = \exp(\mu/T) zV$.

\subsection{van der Waals excluded-volume model }
In the van der Waals excluded-volume model (vdW-EV) the
volume $V$ is substituted by the available
volume $V_{\rm av} = V - v N$, where $v = 16 \pi r^3 / 3$ is the eigenvolume parameter and $r$ is the effective hadron radius parameter.
This results in the following GCE partition function
\eq{\label{evmsingle}
\mathcal{Z}_{\rm EV}(V,T,\mu)
= \sum_{N=0}^{\infty}
\exp\left(\frac{\mu N}{T}\right)
\frac{(V~-~v N)^N}{N!}\,\theta(V-v N)z^N~,
}
where
the $\theta$-function ensures that the sum of eigenvolumes of the particles
does not exceed the total system volume.
In the thermodynamic limit, i.e. when $V \to \infty$, the system pressure is calculated as~\cite{Rischke1991}
\eq{\label{Pex1}
P(T, \mu)~\equiv~
T \frac{\partial \ln \mathcal{Z}_{\rm EV}}{\partial V}
\stackrel{V \to \infty}{\simeq} \frac{T}{V}\,\ln \mathcal{Z}_{\rm EV} 
~=~ P^{\rm id} (T, \mu^*)~,~~~~~~\mu^* = \mu - v \, P(T, \mu)~,
}
where
$P^{\rm id} (T, \mu)=T\,n^{\rm id}(T,\mu)=\exp\left(\mu/T\right)\,T\,z$
is the GCE pressure of the ideal
gas, and $n^{\rm id}$ is the ideal gas particle number density.
The particle number density in the EV model can be calculated as
\eq{\label{nex1}
n(T, \mu) ~\equiv ~ \left(\frac{\partial P}{\partial \mu}\right)_T ~=~ \frac{n^{\rm id} (T, \mu^*)}{1~ + ~v\, n^{\rm id} (T, \mu^*)}.
}

In the GCE one finds that the particle number $N$ fluctuates around its average value $\langle N\rangle=Vn$.
A useful
measure of the particle number fluctuations is the scaled variance $\omega[N]$. It was calculated
analytically in Ref.~\cite{CE-fluc-1}:
\eq{\label{omega}
\omega[N] \equiv \frac{\langle N^2 \rangle - \langle N \rangle^2}{\langle N \rangle}~ =~
(1~-~v\,n)^2~,
}
see also Ref.~\cite{CE-fluc-2}.
Note that analytical expressions in Eqs.~(\ref{Pex1})-(\ref{omega}) are obtained in the
thermodynamic limit $V\rightarrow \infty$. At $v=0$ they are reduced
to the ideal gas expressions. In particular, the particle number distribution ${\cal P}(N)$
is transformed to the Poisson distribution with $\omega[N]=1$.

\subsection{Carnahan-Starling Model}
One can go beyond the standard vdW-EV procedure.  
The Carnahan-Starling (CS) model~\cite{CarnahanStarling} leads
to a better
consistency with the equation of state for classical system of hard spheres.
This model has recently
been applied to hadronic systems~\cite{mf-2014,SBM,Vovchenko:2017cbu}.
The GCE partition function in the CS model can be written as
\eq{
\mathcal{Z}_{\rm CS}(V,T,\mu) ~=~ \sum_{N=0}^{\infty}
\exp\left(\frac{\mu N}{T}\right)~
\exp\left(-~\frac{(4-3\eta) \eta}{(1-\eta)^2} N\right)~\frac{(Vz)^N}{N!}\,\theta\left(V- \frac{1}{4} \,v N \right)~,
}
where $\eta = v \, N / (4V)$ is the packing fraction.

In the thermodynamic limit the pressure is the following
\eq{\label{eq:Pex2n}
P(T,n) = T \, n \, \frac{1+\eta+\eta^2-\eta^3}{(1-\eta)^3}.
}

The GCE particle number density $n(T,\mu)$ and scaled variance $\omega[N]$
can be calculated in the framework of the thermodynamic
mean field approach~\cite{mf-1992,mf-1995,mf-2014}
\eq{\label{CSn}
n(T,\mu)~&=~n^{\rm id}\left[T,\mu~-~T\left(\frac{3-\eta}{(1-\eta)^3}-3\right)\right]~,\\
\omega[N]~& \equiv \frac{T}{n} \, \left(\frac{\partial n}{\partial \mu}\right)_T ~ =~ \frac{(1 - \eta)^4}{(1-\eta)^4 + 8 \, \eta \, (1-\eta/4)}~.
}
In what follows
we consider both the CS and vdW-EV approaches in order to demonstrate
the flexibility of our MC procedure with regards to the variations 
in the EV mechanism.

\subsection{Diagonal Eigenvolume Model}
The single-component vdW-EV model was generalized
to the multi-component case in Ref.~\cite{Yen1997}. It was assumed that the
available volume is the same for  each  hadron  and equals
to the total volume minus the sum of the eigenvolumes of all hadrons in the system.
The GCE partition function has then the following form
for the $f$ hadron species $(i,j=1,\ldots,f)$:
\eq{\label{evm}
\mathcal{Z}_{\rm DE}(V,T,\mu_1,\ldots, \mu_f) = 
\sum_{N_1=0}^{\infty} \ldots \sum_{N_f=0}^{\infty} \, 
 \prod_{i=1}^f
\exp\left(\frac{\mu_i N_i}{T}\right)~
\frac{[(V~-~\sum_j v_j N_j)\,z_i]^{N_i}}{N_i!}\,\theta(V-\sum_{j=1} v_j  N_j)~.
}
We refer to Eq.~(\ref{evm}) as the diagonal eigenvolume (DE) model.
It gives a simple expression for the pressure as
a function of temperature and hadron densities in the thermodynamic limit
\eq{\label{eq:Pex1n}
P(T,n_1,\ldots, n_f) ~= ~T \sum_{i=1}^f \frac{n_i}{1~ -~ \sum_j v_j n_j}~,
}
where the sums go over all types of particles included in the model,
and where
$v_i = 16 \pi r_i^3 / 3$.
In the GCE one has to solve one non-linear equation for the pressure,
\eq{\label{eq:Pex1}
P(T, \mu_1,\ldots,\mu_f) ~= ~\sum_{i=1}^f \, P^{\rm id}_i (T, \mu_i^*)~,~~~~~~ \mu_i^* = \mu_i - v_i \, P(T, \mu)~.
}
The GCE number densities are then calculated as
\eq{\label{eq:nex1}
n_i(T, \mu_1,\ldots,\mu_f) ~=~ \frac{n_i^{\rm id} (T, \mu_i^*)}{1 ~+~ \sum_j v_j \,n_j^{\rm id} (T, \mu_j^*)}~.
}

The DE model~(\ref{eq:Pex1n}-\ref{eq:nex1}) is the
most commonly used one in the thermal model analysis.
For $f=1$, the DE model is reduced to the vdW-EV  model and reproduces
correctly the second virial coefficient for
the system of hard spheres.
However, the DE model
does not treat correctly the cross-terms in the virial expansion of the
multi-component gas of hard spheres.

\subsection{Non-diagonal Eigenvolume Model}
\label{subsec-NDE}
In order to get consistency with the virial expansion for a multi-component system of hard spheres
we use
the model proposed in Refs.~\cite{GKK,Vovchenko:2016ebv}.
The GCE partition function in this model reads
\eq{\label{ndevm}
\mathcal{Z}_{\rm NDE}(V,T,\mu_1,\ldots,\mu_f)=
\sum_{N_1=0}^{\infty} \ldots \sum_{N_f=0}^{\infty} \, 
 \prod_{i=1}^f
\exp\left(\frac{\mu_i N_i}{T}\right)~
\frac{[(V-\sum_j \tilde{b}_{ji} N_j)\,z_i]^{N_i}}{N_i!}\,\theta(V-\sum_{j=1} \tilde{b}_{ji} N_j)~,
}
where
\eq{\label{bij}
\tilde{b}_{ij} = \frac{2\,b_{ii}\,b_{ij}}{b_{ii}+b_{jj}}~,~~~~~b_{ij} = \frac{2 \pi}{3} \, (r_i + r_j)^3~,
}
with $b_{ij}$ being the components of the symmetric matrix of the second virial coefficients \cite{LL}.
We refer to the model given by Eqs.~\eqref{ndevm} and \eqref{bij} as the non-diagonal eigenvolume (NDE) model\footnote{In Ref.~\cite{Vovchenko:2016ebv} it is called the ``Crossterms'' EV model.}.

The pressure of the NDE model has the following form in the thermodynamic limit
\eq{\label{eq:Pex2n}
P(T,n_1,\ldots, n_f) = \sum_{i=1}^f P_i = T \sum_{i=1}^f \frac{n_i}{1 - \sum_j \tilde{b}_{ji} n_j}~,
}
where the $P_i$ quantities  can be regarded as ``partial'' pressures.
In the GCE formulation one has to solve the following system of non-linear
equations for $P_i$:
\eq{\label{eq:Pex2pi}
P_i = P_i^{\rm id} \left(T, \mu_i - \sum_{j=1}^f \tilde{b}_{ij} \, P_j \right), \qquad i = 1, \ldots , f,
}
Hadronic GCE densities $n_i$ can then be recovered by solving the system of linear equations connecting $n_i$ and $P_i$:
\eq{\label{eq:Pex2ni}
T n_i + P_i \sum_{j=1}^f \tilde{b}_{ji} n_j = P_i, \qquad i = 1, \ldots , f~.
}

\section{
Canonical Ensemble}\label{sec-CE}
In the CE,
the conserved charges
are conserved in each microscopic state of the system.
This can be achieved by adding the corresponding Kronecker delta functions in the GCE
partition function.
For the four EV models described in the previous section
one has the following CE partition functions:
\eq{
Z_{\rm EV}(V,T,N) ~&=~ \frac{(V~-~v N)^N}{N!}\,
z^N~\theta(V - vN), \label{CE-EV}\\
%}
Z_{\rm CS}(V,T,N)  ~&=~ \frac{1}{N!} \, \left[ zV \, \exp\left(-~\frac{(4-3\eta) \eta}{(1-\eta)^2}\right) \right]^N ~ \theta\left(V- \frac{1}{4} \,v N \right),
\label{CE-CS} \\
%}
\label{eq:DiagCE}
 Z_{\rm DE}(V,T,\{Q\})  ~& =~\sum_{N_1=0}^{\infty} \ldots \sum_{N_f=0}^{\infty} \, 
 \prod_{i=1}^f \, \frac{\left[ (V - \sum_j v_j N_j) \,  z_i
\right]^{N_i}}{N_i!}\,  \nonumber\\
& \times ~\theta(V - \sum_j v_j N_j)\,  \prod_{k=1}^c \delta(Q_k - \sum_j q_k^{(j)} N_j)~,\\
%}
\label{eq:CrossCE}
 Z_{\rm NDE}(V,T,\{Q\})  ~ &=~ \sum_{N_1=0}^{\infty} \ldots \sum_{N_f=0}^{\infty} \, 
 \prod_{i=1}^f \,
 \frac{\left[ (V - \sum_j \tilde{b}_{ji} N_j) \,  z_i
\right]^{N_i}}{N_i!}\, \nonumber \\
& \times ~\theta(V - \sum_j \tilde{b}_{ji} N_j)\, \prod_{k=1}^c \delta(Q_k - \sum_j q_k^{(j)} N_j)~.
}
In Eqs.~(\ref{eq:DiagCE}) and (\ref{eq:CrossCE})
for multi-component systems, $\{Q\} = Q_1,\ldots,Q_c$ are the set of
conserved charges and $q_k^{(j)}$ is the $k$th charge of the
particle species $j$. For a single-component case there we
identify the single conserved charge $Q$ with the particle number $N$, i.e. $Q \equiv N$.

For the ideal gas, i.e. for $v_i \equiv 0$ in \eqref{eq:DiagCE} or $\tilde{b}_{ij} \equiv 0$ in \eqref{eq:CrossCE},
the thermodynamic properties can be calculated analytically~\cite{Begun2004}.
To our knowledge, no approach has been developed to calculate the moments of the multiplicity
distribution for
the EV models in the CE formulation of HRG.

\section{Monte Carlo approach}\label{sec-MC}

\subsection{Grand Canonical Ensemble}

The GCE partition functions listed in Sec.~\ref{sec-an} determine the probability distribution of particle numbers at given values of the thermodynamic parameters for the corresponding excluded volume  models.
In most general case, the probability o having a microstate
with a set of particles numbers
$(N_1, \ldots, N_f)$
has the form
\eq{\label{prob}
{\cal P}(N_1, \ldots, N_f;V,T,\{\mu_Q\})  \propto  F(N_1, \ldots, N_f;V,T,\{\mu_Q\}) \, \times \, \Theta(N_1, \ldots, N_f;V),
}
where $\Theta(N_1, \ldots, N_f;V)$
ensures that only the microstates where the sum of all proper
particle eigenvolumes does not exceed the total volume of the system are considered, and
$\{\mu_Q\}\equiv \mu_1,\ldots , \mu_c$ corresponds to the independent chemical potentials
which
regulate the conserved charges $Q_1,\ldots,Q_c$ in the system.
The function $F(N_1, \ldots, N_f;V,T,\{\mu_Q\})$ is a smooth function of particle numbers
within the domain of allowed microstates. The chemical potential of $i$th particle species is
\eq{\label{mui}
\mu_i~=~\sum_{k=1}^c q_k^{(i)}\,\mu_k~,
}
where $q_k^{(i)}$ is the $k$th charge of the $i$th particle.
In the HRG the number of conserved charges is normally much smaller than the number of particle species
(i.e., $c \ll f$).
It is evident that $F$ is defined up to a multiplicative  factor which may
depend on thermodynamic variables but is independent of the particle numbers.

Both the $F$ and $\Theta$ functions are well defined
for the models
listed in Sec.~\ref{sec-an}:
\eq{
F_{\rm EV}(N;V,T,\mu) & ~=~ \frac{\left[ (V-vN) \,  z
\, e^{\mu/T}
\right]^N}{N!}~, \\
\Theta(N;V) & ~=~ \theta(V - vN)~; \\
%}
\nonumber \\
%}
F_{\rm CS}(N;V,T,\mu) & ~=~ \frac{1}{N!} \, \left[ zV \, \exp\left(-~\frac{(4-3\eta) \eta}{(1-\eta)^2}\right) \,
\, e^{\mu/T}
\right]^N~, \\
\Theta(N;V) & ~=~ \theta\left(V- \frac{1}{4} \,v N \right)~;
}
\eq{
\label{eq:DiagonalMC}
F_{\rm DE}(N_1, \ldots, N_f;V,T,\{\mu_Q\}) & ~=~ \prod_{i=1}^{f} \, \frac{\left[ (V - \sum_j v_j N_j) \,  z_i
\, e^{\mu_i/T}
\right]^{N_i}}{N_i!}~, \\
\Theta(N_1, \ldots, N_f;V) & ~=~ \theta(V - \sum_j v_j N_j)~;\\
%}
\nonumber \\
F_{\rm NDE}(N_1, \ldots, N_f;V,T,\{\mu_Q\}) & ~=~ \prod_{i=1}^{f} \, \frac{\left[ (V - \sum_j \tilde{b}_{ji} N_j) \,  z_i
\, e^{\mu_i/T}
\right]^{N_i}}{N_i!}~, \label{ct-F}\\
\Theta(N_1, \ldots, N_f;V) & ~=~ \prod_{i=1}^{f} \, \theta(V - \sum_j \tilde{b}_{ji} N_j)~. \label{ct-T}
}

In the ideal gas limit
the probability ${\cal P}$ (\ref{prob})
is reduced to a product of the $f$
independent Poisson distributions, i.e. ${\cal P} \propto \Pi$ where
\eq{\label{Pi}
\Pi(\{N_i\};V,T,\{\mu_Q\}) ~=~ \prod_{i=1}^f \, \frac{\langle N_i \rangle^{N_i}}{N_i!}\, e^{-\langle N_i \rangle}~.
}
The probability function ${\cal P}$ (\ref{prob})
cannot be decomposed into a product of independently distributed variables in the presence of finite eigenvolumes in a multi-component system.
Thus, a straightforward sampling of particle numbers looks problematic.
To avoid this problem we rewrite the probability
${\cal P}$ (\ref{prob})
in the following form
\eq{\label{prob-1}
{\cal P}(\{N_i\};V,T,\{\mu_Q\}) \propto \frac{F(\{N_i\};V,T,\{\mu_Q\})}
{\Pi(\{N_i\};V,T,\{\mu_Q\})} \, \times \, \Pi(\{N_i\};V,T,\{\mu_Q\}) \, \times \, \Theta(\{N_i\};V),
}
where $\Pi(\{N_i\};V,T,\{\mu_Q\})$ is an auxiliary function, taken in the form of Eq.~(\ref{Pi})
with Poisson rate parameters $\langle N_i \rangle$ which can, in general,
be chosen arbitrarily and differently for different values of
$V$, $T$, and $\{\mu_Q\}$.
The Monte Carlo (MC) sampling of the particle numbers can be then performed with the help of the
importance sampling technique (see e.g. \cite{sampl}).
In practical calculations, the parameters $\langle N_i \rangle$ should be 
chosen in a way so that the auxiliary distribution $\Pi$ resembles the true 
distribution $F$ as closely as possible. This helps to avoid 
oversampling of the ``unimportant'' low-probability regions and
makes the statistical convergence faster. 
In our calculations we will utilize the multi-Poisson distribution 
in Eq.~\eqref{Pi} with parameters $\langle N_i \rangle$ calculated 
within the corresponding analytic models defined in Sec.~\ref{sec-an}.
Of course, it is also possible to use an auxiliary distribution which is different from the multi-Poisson distribution in Eq.~\eqref{Pi}, especially if it improves the statistical convergence.

Denoting the ratio $F / \Pi$ as a weight $w$,
the probability distribution can be written
\eq{\label{prob-2}
P(\{N_i\};V,T,\{\mu_Q\}) \propto w(\{N_i\};V,T,\{\mu_Q\}) \, \times \,
\Pi(\{N_i\};V,T,\{\mu_Q\}) \, \times \, \Theta(\{N_i\};V)~.
}
The MC sampling procedure includes the following steps:
\begin{enumerate}
\item Sample the numbers $(N_1, \ldots, N_f)$ from the auxiliary multi-Poisson distribution $\Pi$ (\ref{Pi}).
\item If the indicator function $\Theta$~\eqref{prob} fails for the sampled numbers, then reject the event and go back to the first step. If $\Theta$ passes, then go to the next step.
\item Calculate the weight $w=F/\Pi$ and accept the event with this weight.
\item Go back to step 1 to generate a new event, or terminate the procedure if the desired number of the generated events is achieved.
\end{enumerate}

Let us have  $l=1,\ldots,M$ samples of particle numbers $\{N_i\}_l$ with weights $w_l$. The sample mean of any function $f(N_1,\ldots,N_f)$ of the particle numbers is calculated in the following way
\eq{
\langle f(N_1,\ldots,N_f) \rangle_M~ = ~\frac{\sum_{l=1}^M w_l f(\{N_i\}_l)}{\sum_{l=1}^M w_l}.
}
In the limit $M \to \infty$ the sample mean will converge to the GCE expectation value, i.e.
\eq{
\langle f(N_1,\ldots,N_f) \rangle_M ~\xrightarrow[M \to \infty]{}~ \langle f(N_1,\ldots,N_f) \rangle_{\rm GCE}~.
}

The statistical error estimate for $\langle f(N_1,\ldots,N_f) \rangle_M$ reads 
\eq{
\sigma^2_{f} ~ = ~\frac{\sum_{l=1}^M w_l^2 [ f(\{N_i\}_l) - \langle f \rangle_M ]^2}{(\sum_{l=1}^M w_l)^2}.
}

\subsection{Monte Carlo Method in the Canonical Ensemble}
\label{subsec-CE}
In the CE, the conserved charges $\{Q\} = Q_1,\ldots,Q_c$
in the system are fixed to their exact values in each microscopic state.
The exact charge conservation is enforced by adding the corresponding Kronecker
delta functions into the probability distribution, i.e.
\eq{\label{prob-CE}
{\cal P}(\{N_i\};V,T,\{Q\})  \propto
F(\{N_i\};V,T,\{\mu_Q=0\})  \times  \Theta(\{N_i\};V)
\times  \prod_{k=1}^c \delta(Q_k - \sum_j q_k^{(j)} N_j)\,.
}
Similarly to the GCE, the MC approach within the CE
proceeds by introducing the product of auxiliary Poisson distributions, i.e.
\eq{\label{prob-CE1}
{\cal P}(\{N_i\};V,T) & ~\propto ~ w(\{N_i\};V,T,\{Q\}) \, \times \, \Pi(\{N_i\};V,T,\{Q\}) \nonumber \\
 & ~\times \, \Theta(\{N_i\};V) \, \times \, \prod_{k=1}^c \delta(Q_k - \sum_j q_k^{(j)} N_j)~.
}
The MC sampling in the CE includes only one additional step compared to the corresponding procedure in the GCE:
if the generated configuration does not satisfy the exact charge conservation laws then it is rejected.
Our approach is quite similar to the importance sampling in an ideal HRG in the micro canonical ensemble performed previously in~Refs.~\cite{BecattiniMCE1,BecattiniMCE2}.

It should be noted that a naive, straightforward implementation of rejection sampling described above would be rather inefficient and time-consuming, as 
the probability to choose a set of random charges satisfying conservation laws is very small.
We, therefore, use the multi-step procedure of~Ref.~\cite{BecattiniMCE2} for sampling particle yields in the CE.
In this procedure one first separately generates the total number of baryons and antibaryons from the Poisson distribution.
If the generated net baryon number does not satisfy the baryon number conservation then the configuration is rejected outright, without performing the time-consuming generation of all the individual hadron yields.
If the baryon number conservation is fulfilled, then the numbers of all individual (anti)baryons are sampled from the multinomial distribution, and the whole procedure is repeated in the same fashion for (anti)strange mesons, and then for the remaining (anti)charged mesons.
The fact that most of the unsuitable configurations are rejected at an early step in this procedure
gives a significant performance boost as compared to the straightforward independent sampling of all particle multiplicities from a multi-Poisson distribution.

The procedure described above can also be applied to a HRG with van der Waals interactions~\cite{Vovchenko:2016rkn,Vovchenko:2017zpj}.
This model contains, in addition to excluded volume effects, the attractive interactions between hadrons in the mean-field approximation.
The details of the corresponding MC procedure are given in the Appendix.

\section{Calculation results}\label{sec-calc}

\subsection{Finite-size Effects in the Grand Canonical Ensemble}
Let us consider first a single-component gas with 
EV interactions in the GCE in the vdW-EV model.
When the EV effects are present,
the intensive quantities
depend explicitly on the total system volume.
Most notably, the particle density
equals zero if the system volume $V$ is smaller than the eigenvolume of a single particle.
The finite-size effects cannot be described by the analytic formulas presented in Sec.~\ref{sec-an},
as they all are derived under the assumption of the thermodynamic limit.
However, these effects can be studied with the help of the MC procedure described in Sec.~\ref{sec-MC}.

We consider a simple example to illustrate the finite-size effect.
We assume a single-component gas of particles with the mass of 1~GeV, which is a typical energy scale for hadronic systems.
We consider the vanishing chemical potential, i.e., $\mu=0$, and a temperature of $T=150$~MeV.
In order to mimic the presence of large number of hadron states in a realistic HRG we use a rather high degeneracy factor of $g = 150$ in our calculation.
This is important as the magnitude of the eigenvolume effects scales with the
total number of the finite-sized hadrons
in the system.

\begin{figure}[t]
\centering
\includegraphics[width=0.75\textwidth]{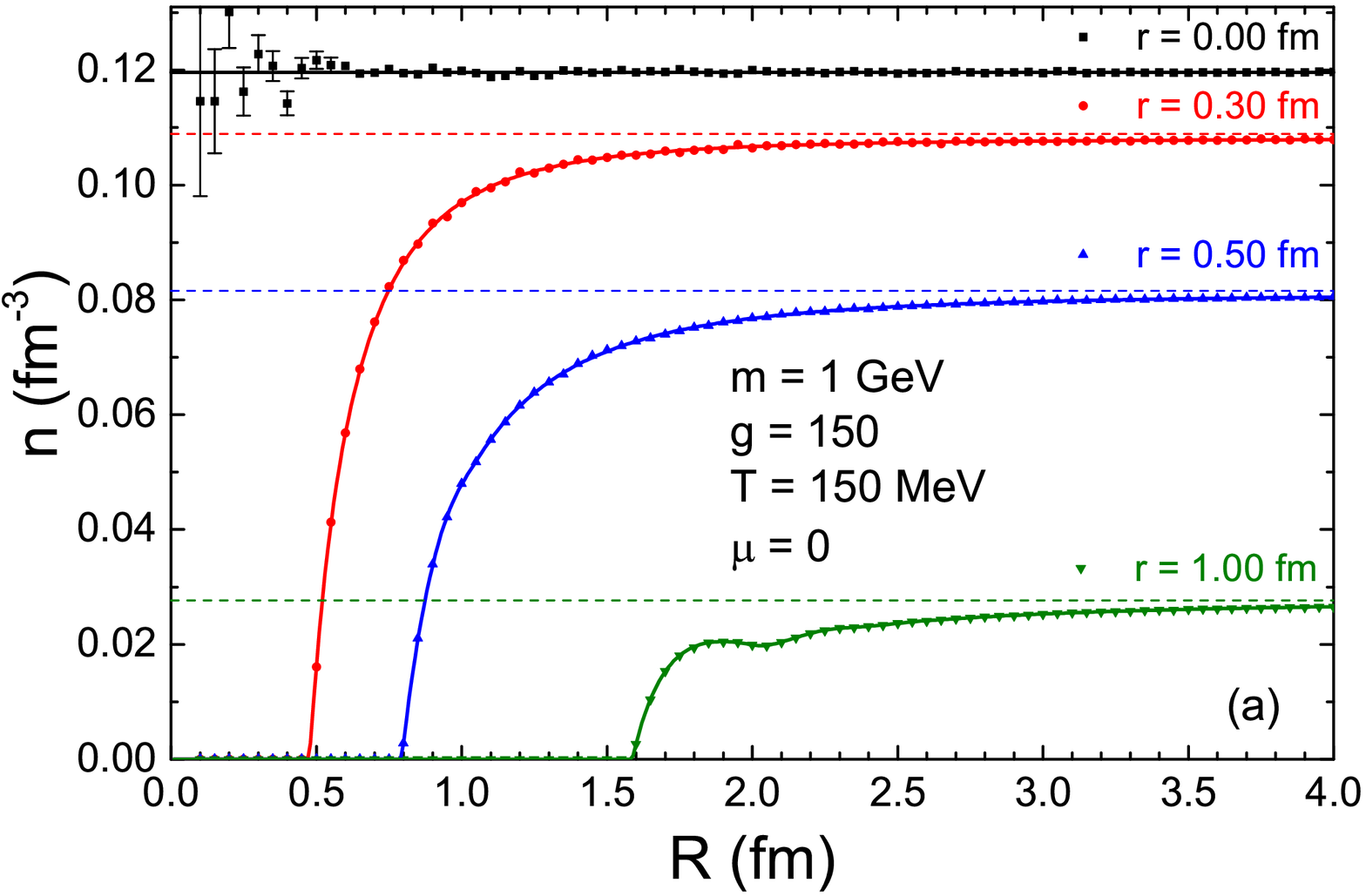}
\includegraphics[width=0.75\textwidth]{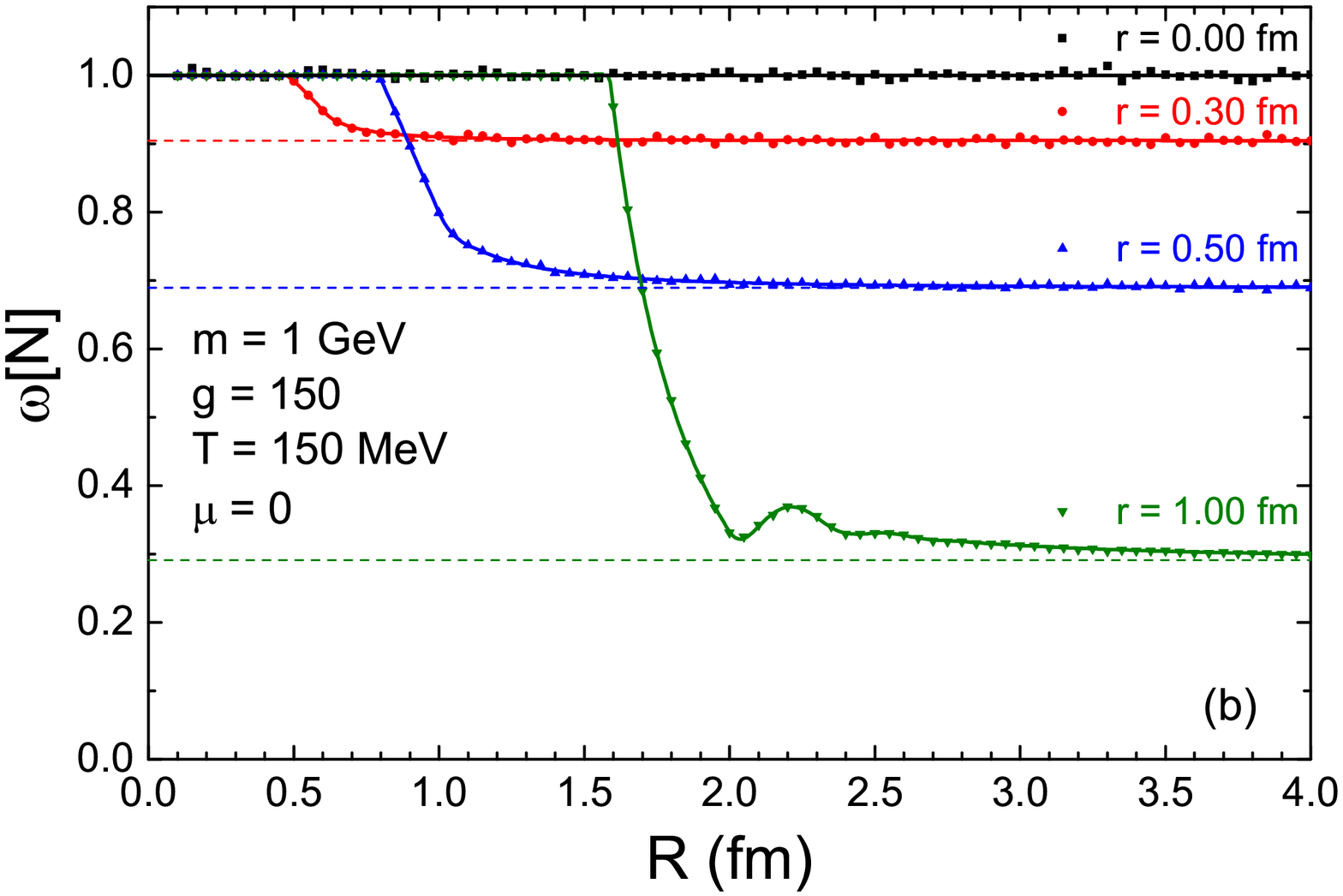}
\caption[]{
(a) The GCE particle number density $n$ and (b) the scaled variance $\omega[N]$
as functions on the system radius $R$ for the EV model for particles of mass $m=1$~GeV
and degeneracy $g=150$ at $T=150$~MeV and $\mu=0$.
Dots show the MC results for four different values of the hard-core radius: $r$ = 0, 0.3~fm, 0.5~fm, and 1~fm.
Dashed horizontal lines show the values of the particle density (a) and scaled variance (b) calculated in the
thermodynamic limit from Eqs.~(\ref{nex1}) and (\ref{omega}), respectively.
Solid lines show the analytic results obtained by the direct summation of the GCE partition function.
}\label{fig:FSS}
\end{figure}

The system-size dependence of the particle 
number density, $n = \langle N \rangle / V$, is calculated using the MC method.
Additionally, we consider the scaled variance, $\omega[N]$, of the particle number fluctuations.
The Poisson rate parameter $\langle N \rangle$ in the auxiliary distribution $\Pi$ (\ref{Pi})
is taken to be $\langle N \rangle = n_{\rm EV}(T,\mu=0;r) \, V$,
where $n_{\rm EV}(T,\mu=0;r)$ is the particle number density in the thermodynamic limit ($V \to \infty$), calculated analytically using Eqs.~(\ref{Pex1}) and (\ref{nex1}).
The dependence of $n$ on the total system radius $R$ (defined as $V \equiv 4\pi R^3 / 3$)
is depicted in Fig.~\ref{fig:FSS}
for four different values of the effective particle radius parameter ($r = 0,\,0.3$, 0.5, and 1~fm).
For each pair of the $R$ and $r$ values we generate and perform an averaging over $10^5$ MC events.
The calculations show a consistent approach of the particle density $n$ to its limiting value
with increasing $R$.
The resulting limiting values at large $R$ in all cases appear to coincide with the corresponding values in the thermodynamic limit calculated from Eqs.~(\ref{nex1}) and (\ref{omega}). This is an expected result.

The number of terms in the GCE EV partition function~\eqref{evmsingle} is
finite due to the presence of the $\theta$-function.
Thus, it is also possible
to calculate the moments of the multiplicity distribution analytically,
by explicitly summing over all $N$-states. 
More specifically, the GCE average of arbitrary function $f(N)$ of the particle number is
calculated as
\eq{\label{eq:MC:analyt}
\langle f(N) \rangle = \frac{\displaystyle \sum_{N=0}^{\lfloor V/v \rfloor} \, f(N) \, Z_{\rm EV}(T,V,N)}{\displaystyle \sum_{N=0}^{\lfloor V/v \rfloor} \, Z_{\rm EV}(T,V,N)}.
}
We have performed such a calculation
in order to cross-check our MC results. The results of these analytic calculations
are shown in Fig.~\ref{fig:FSS} by solid lines and they are fully consistent with the MC results.
Note that a calculation of a direct sum over all states in the grand canonical
partition function becomes numerically intractable in the multi-component gas with a large number of components. 
The MC procedure, on the other hand, does not suffer from such a complication.

As seen from Fig.~\ref{fig:FSS}a, both the analytical and the MC calculations exhibit the presence of a small region where the particle number density locally decreases with an increase of the system volume for $r = 1$~fm. 
A pronounced presence of such region(s) was also verified for larger values of particle radius parameter $r$.
This result seems counterintuitive.
Recall, however, that the particle density is given as the ratio $n \equiv \langle N \rangle / V$.
The number of terms $N_{\rm tot} = \lfloor V/v \rfloor$ in Eq.~\eqref{eq:MC:analyt}, which is used to calculate $\langle N \rangle$, is finite. 
The $N_{\rm tot}$ increases by one once the ratio $V/v$ reaches the next integer number.
However, until that happens, the $N_{\rm tot}$ value is fixed and this severely limits the growth of $\langle N \rangle$ with $V$.
For this reason, the ratio $n = \langle N \rangle / V$ can locally be a decreasing function of $V$.
The same mechanism is responsible for appearance of non-monotonous regions in the $V$-dependence of the scaled variance, $\omega[N]$, seen in Fig.~\ref{fig:FSS}b.
On the other hand, the dependence of $\langle N \rangle$ on $V$ remains strictly monotonically increasing in all cases.

The nonmonotonic system volume dependence of the particle number density appears in the model for small systems, when the eigenvolume of a single particle is not negligible compared to the system volume, and when the EV effects are strong.
The appearance of non-monotonicities with respect to the overall system size was also reported in the microcanonical ensemble calculation in Ref.~\cite{Keranen:2004eu}, where the effect was associated with the proximity to the production energy threshold.
Thus, the non-monotonic behavior of thermodynamic observables might be a generic feature of small systems, where the size or energy of a single constituent particle is non-negligible compared to the total system size or energy.
It would be interesting to consider these effects in real physical systems, not necessarily those created in high-energy collisions.

\begin{figure}[t]
\centering
\includegraphics[width=0.8\textwidth]{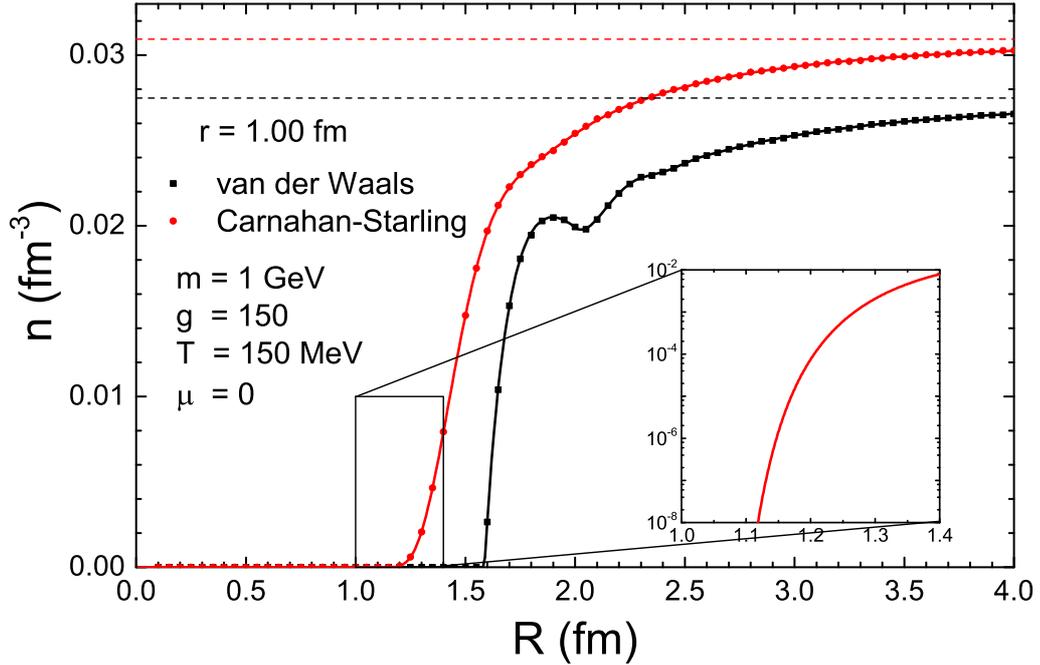}
\caption[]{
The dependence of the GCE particle number density $n$ on the total system
radius $R$ calculated within the EV (black) and Carnahan-Starling (red) models for hard-core radius of $r = 1$~fm.
The solid lines show the analytic results obtained by the direct summation of the grand canonical partition function.
All system parameters are the same as for calculations shown in Fig.~\ref{fig:FSS}.
The inset shows the analytic Carnahan-Starling model calculations in the vicinity of the threshold system radius $R = 1$~fm, on the logarithmic scale.
}\label{fig:VDW-vs-CS}
\end{figure}

The MC procedure is quite flexible to the variations in the excluded volume mechanism used.
We perform the calculations for the CS model in order to illustrate this fact.
The dependence of the particle number density on the system radius $R$ is shown in Fig.~\ref{fig:VDW-vs-CS}.
A difference between the EV and the CS models is most significant for large values
of particle radius parameter $r$ and/or at high particle number densities. Thus, we only show the results for the case $r=1$~fm.
In the CS model, the particle number density $n$ approaches from below
the corresponding limiting value (\ref{CSn}) with increasing system size $R$.
The calculations also show that CS values of $n$ are generally larger then the EV ones at all values of $R$.

It is evident that there exists a minimum system volume, characterized by the system radius $R_{\rm min}$, such that the particle number density is strictly zero for $R < R_{\rm min}$.
In the van der Waals EV model one has $R_{\rm min} = 4^{1/3} r$.
For $r = 1$ fm one obtains $R_{\rm min} \simeq 1.59$~fm, 
the calculations in Fig.~\ref{fig:VDW-vs-CS} are consistent with this expectation.
For the CS model one has a smaller value of the minimum system radius, $R_{\rm min} = r$.
However, there is a very strong suppression of the particle number density for system volumes which are only slightly larger than the minimum system volume in the CS model, this fact is illustrated in the inset of Fig.~\ref{fig:VDW-vs-CS}.

\subsection{Simultaneous Effects of Canonical Ensemble and Excluded-Volume}

\begin{figure}[t]
\centering
\includegraphics[width=0.75\textwidth]{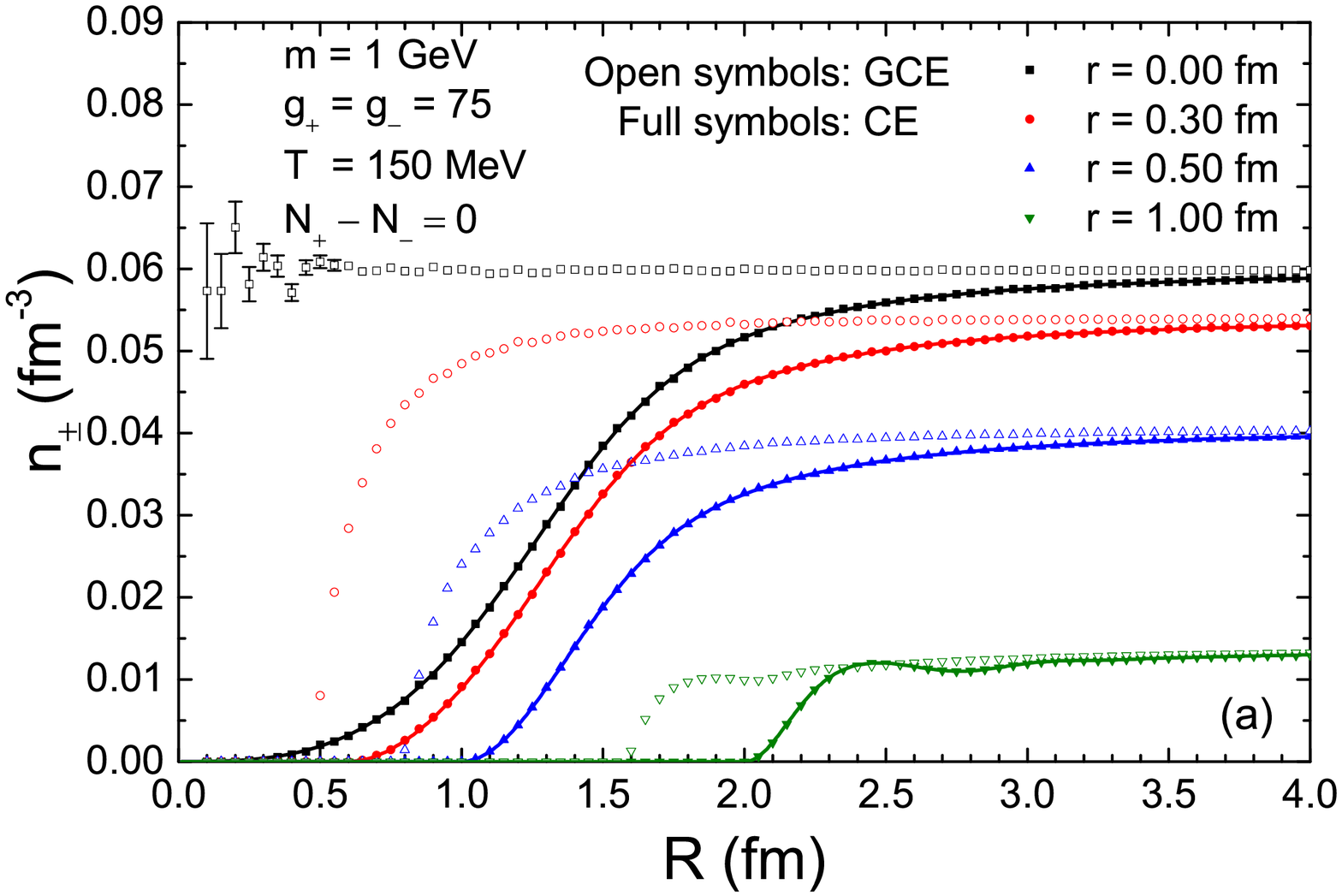}
\includegraphics[width=0.75\textwidth]{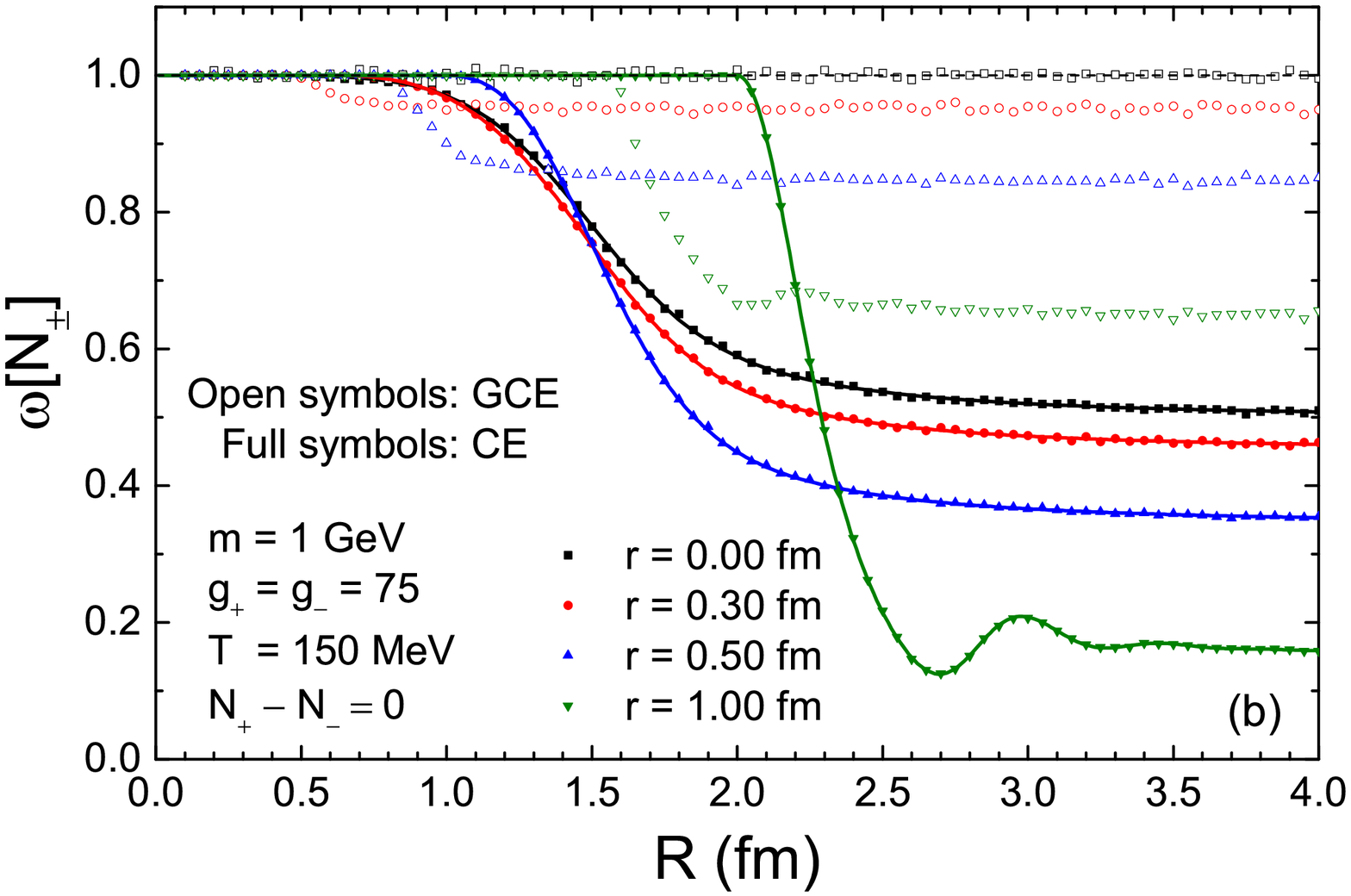}
\caption[]{
The MC results for (a) $n_{\pm}$ and (b) $\omega[N_{\pm}]$
as functions of $R$. The MC calculations are performed at
$m=1$~GeV, $g_{\pm}=75$, $T=150$~MeV, and $Q\equiv N_+ -  N_-=0$.
Open symbols show the MC results in the GCE
and full symbols in the CE
for four different values of hard-core radius: $r = 0$, 0.3~fm, 0.5~fm, and 1~fm.
Solid lines
show the analytic results obtained by the direct summation of the  partition function.
The lines for $r=0$ coincide with values (a) $n_\pm$ and (b) $\omega[N_\pm]$
calculated analytically in Ref.~\cite{Begun2004}.
}\label{fig:FSS-CE}
\end{figure}

In order to study the excluded-volume effects in the CE
we consider a two-component system of particles and antiparticles.
The degeneracy factor of $g_{\pm} = 75$, the particle mass of $m_\pm=1$~GeV,
zero net charge, $Q=N_+-N_-=0$, and the system temperature $T=150$~MeV are employed.
Using the MC method we calculate the system-size dependence
of the (anti)particle number density $n_{\pm}$ and the scaled variance $\omega[N_{\pm}]$.
The MC CE and the MC GCE results
for four different values of
the particle radius parameter
are shown in Fig.~\ref{fig:FSS-CE}.

The MC results for $n_{\pm}$ and $\omega[N_\pm]$ at $r=0$ can be directly
compared to the analytical results for the ideal gas
obtained in Ref.~\cite{Begun2004}.
Our MC calculations are fully consistent with these analytical
results (shown by black solid lines). 
In particular, $\omega[N_\pm]=1/2$ at $R\rightarrow \infty$ for the CE.
It is notable that while $\omega[N_\pm] \to 1/2$ in the CE, one has $\omega[N_\pm] \to 1$ as volume goes to infinity.
This means that $\omega[N_\pm]$ has different limiting values between the CE and the GCE in the thermodynamic limit, in contrast to the particle number densities~(Fig.~\ref{fig:FSS-CE}a), which tends to the same limit in both ensembles.
This difference may seem counterintuitive in light of the expected thermodynamic equivalence of different ensembles in the infinite volume limit.
Recall, however, that the thermodynamic equivalence of ensembles extends to mean values, but not to fluctuations, hence the observed difference between the CE and the GCE. We refer to Ref.~\cite{Begun2004} where this question was studied in great detail in an analytic model.

The analytic results for $r>0$, obtained from a direct summation of the partition function, are also
shown in Fig.~\ref{fig:FSS-CE} by the colored solid lines. They are fully consistent with the MC results.
The presence of the CE effects due to the exact charge conservation leads
to a further suppression of $n_\pm$
at a finite $R$,
in addition to the suppression resulting from the
EV effects.
The same is generally true for $\omega[N_\pm]$.
There is, however, one important difference.
The CE suppression effects for $n_\pm$
disappear
in the thermodynamic limit $R\rightarrow \infty$ and only the EV suppression effects remain, whereas
both the CE and the EV suppression effects
for $\omega[N_\pm]$ survive.
In particular, at $R\rightarrow \infty$ the CE values of $\omega[N_\pm]$
shown in Fig.~\ref{fig:FSS-CE}b
are smaller at $r>0$ than the ideal gas CE value of
1/2. At $R\rightarrow \infty$, the CE values of $\omega[N\pm]$
are also smaller than the corresponding GCE limiting values at the same $r$ shown in Fig.~\ref{fig:FSS}b.

It is seen from Fig.~\ref{fig:FSS-CE}a that there is a minimum system volume, below which the 
particle number density is strictly zero, similar to the GCE case.
However, this minimum volume is approximately twice larger in the CE as compared to the GCE.
The reason is that no microstate with a single particle is permitted in the CE since that would violate the exact charge neutrality condition.
The presence of an antiparticle for each particle is required.
Therefore, the minimum system volume has to accommodate at least two particles with a finite eigenvolume.

\subsection{Hadron number fluctuations in HRG}
The MC formulation of
the full HRG model can be used to describe the hadron yields and their fluctuations in the presence of both the EV interactions and the exact charge conservation effects.

\begin{figure}[t]
\centering
\includegraphics[width=0.75\textwidth]{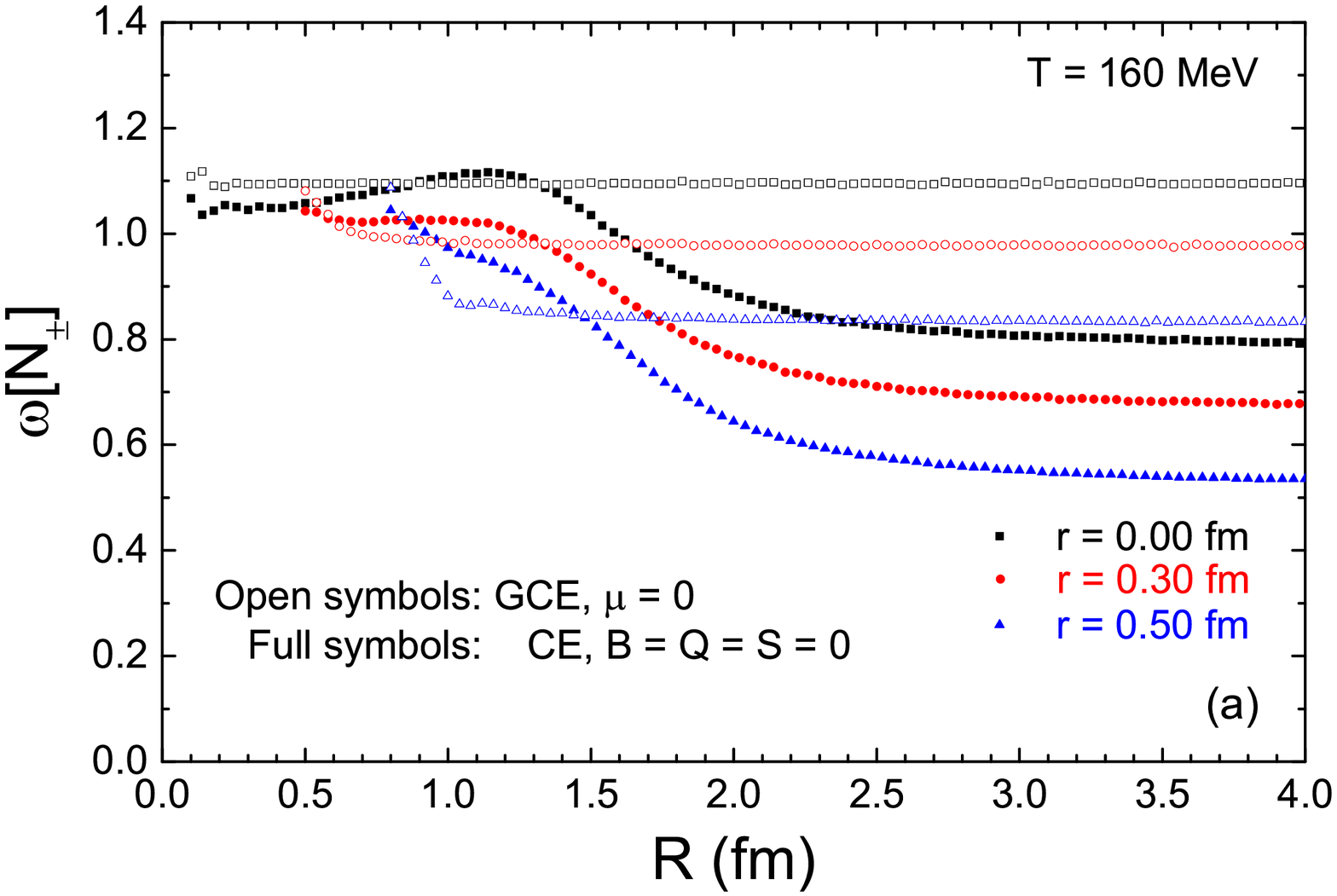}
\includegraphics[width=0.75\textwidth]{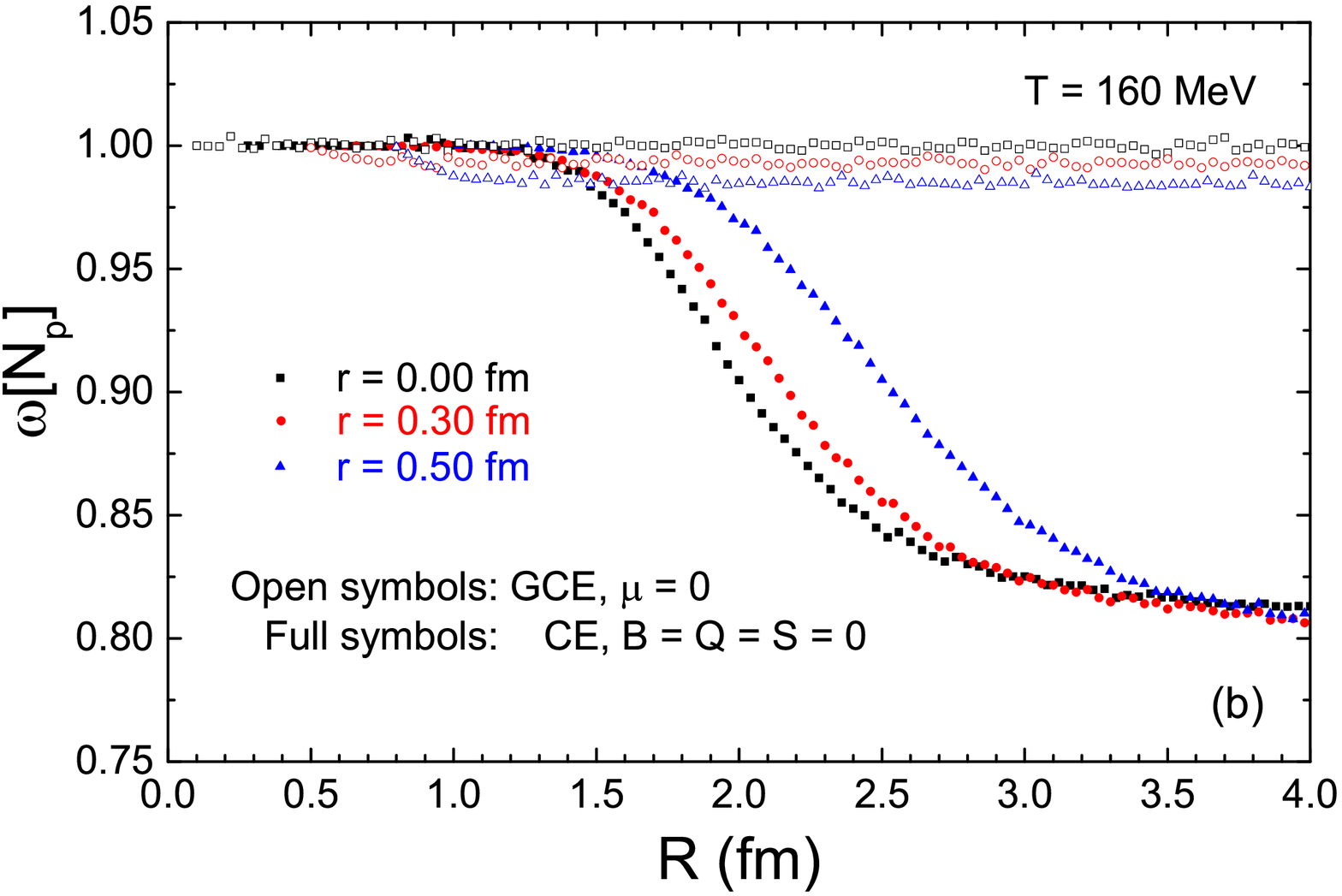}
\caption[]{
The MC results for (a) $\omega[N_{\pm}]$ and (b) $\omega[N_p]$
in full HRG at $T=160$~MeV
as functions of $R$.
Open symbols show the MC results in the GCE with $\mu_B = \mu_Q = \mu_S = 0$
while full symbols depict the MC results in the CE for the $B = Q = S = 0$ system.
Three different values of hard-core radius are considered: $r = 0$, $0.3$~fm and $0.5$~fm.
}\label{fig:FSS-HRG}
\end{figure}

To illustrate the role of both EV and exact charge conservation effects in HRG
a system with zero conserved charges, $B=S=Q=0$, will be considered at first. It may correspond to hadron
states
created in $p\overline{p}$ or $e^+e^-$ reactions.
Three values of the hadron hard-core radius, $r = 0$ (ideal HRG), 0.3, and 0.5~fm,
the same for all hadron species, are considered within the MC formulation of the EV HRG, containing 361 different hadron species. 
We apply here the diagonal EV model formulation, one should note here, however, that in the considered case of equal radii for all hadron species,
both the diagonal and the non-diagonal EV models are equivalent.
In Fig.~\ref{fig:FSS-HRG}a the scaled variance $\omega[N_\pm]$ of the number of all positively or negatively charged hadrons in HRG
is shown as a function of the system radius $R$. The system temperature is fixed at $T=160$~MeV.
In these calculations we additionally take into account contributions to $N_\pm$ from resonance decays.
Thus, the MC procedure contains one additional step at the end: simulation of the chain of probabilistic decays of all resonances.

From Fig.~\ref{fig:FSS-HRG}a one observes that both EV and exact charge conservation
effects suppress the $N_\pm$ fluctuations in the thermodynamical limit $R\rightarrow\infty$. For
$r=0.5$~fm the numerical values of both suppression effects are rather similar. At small $R$ the fluctuations
are additionally sensitive to the finite size effects.

In Fig.~\ref{fig:FSS-HRG}b the scaled variance $\omega[N_p]$ for the fluctuations of the number of protons
is shown as a function of the system radius $R$. 
Here the EV effects are defined by the total number $N^{\rm prim}_{\rm tot}$ of primary hadrons and resonances.
The mean number of protons $\langle N_p\rangle$ is suppressed significantly by the presence
of the excluded volume $vN_{\rm tot}^{\rm prim}$. 
However, as $\langle N_p\rangle$ is much smaller than
$\langle N^{\rm prim}_{\rm tot}\rangle$,  the fluctuations of
$N_p$ have only a minor influence on the event-by-event values of the total excluded volume.
The magnitude of the EV effects on $\omega[N_p]$ scales approximately as $\langle N_p \rangle / \langle N^{\rm prim}_{\rm tot} \rangle$~(see Ref.~\cite{CE-fluc-1}), and the $N_p$ fluctuations therefore do not deviate significantly from Poisson distribution.
This is not the case for the $N_\pm$ fluctuations, as $\langle N_\pm\rangle$
is comparable with $\langle N^{\rm prim}_{\rm tot}\rangle$. 
We do note that higher order proton number fluctuations, not considered in this work,
are more sensitive to both excluded-volume~\cite{Vovchenko:2017ayq} and exact charge conservation~\cite{Bzdak:2012an} effects.

The CE suppression effects for $\omega[N_p]$ survive in the thermodynamic limit $R\rightarrow\infty$, and they are not very sensitive to the value of the radius parameter $r$. 
The main source of the CE suppression of $\omega[N_p]$ is the exact conservation of the net baryon number $B=0$.
It is interesting that the approach of $\omega[N_p]$ to the CE thermodynamic limit is slower when a larger value of $r$ is used, as seen in Fig.~\ref{fig:FSS-HRG}(b).
The physical reason is that, at a fixed system volume, the EV effects reduce the number of particles in the system, 
thereby effectively moving the system farther away from the thermodynamic limit.

\subsection{Fit of the Hadron Yields in p+p Collisions}
The CE formulation of HRG can be used to describe the hadron yield data in
collisions of small systems, such as (anti)proton-proton and $e^+\,e^-$ collisions.
Previously, only the non-interacting HRG was used in such
studies~\cite{Becattini1995,Becattini1997,Becattini2000,Becattini2010,VBG2015}.
Here we will demonstrate the effect of the finite hadron eigenvolumes on chemical freeze-out parameters.
For this purpose we analyze the hadron yield data of
the NA61/SHINE Collaboration in inelastic
proton-proton interactions at beam laboratory momentum $p_{\rm lab} = 31,\,40,\,80,\,158$~GeV/$c$~\cite{NA61-p+p-mult}.
The experimental data contains yields\footnote{The newer data at some of the collision energies now also contain the yields of $p$, $\Lambda$, and/or $\phi$. In the present work we retain the same hadron yield dataset which we previously analyzed in Ref.~\cite{VBG2015}.} of $\pi^-$, $\pi^+$, $K^-$, $K^+$, and $\bar{p}$.
These data were recently analyzed in Ref.~\cite{VBG2015} within the ideal HRG in the CE.
It was found that the data can be reasonably well described with three chemical
freeze-out parameters: the temperature $T$, the system radius (volume) $R$,
and the strangeness undersaturation parameter $\gamma_S$.

To illustrate the effect of finite hadron eigenvolumes on chemical
freeze-out parameters let us consider a simple case when all hadrons have the same hard-core radius $r$.
Hadron densities become suppressed compared to the ideal gas.
In the GCE, the suppression factor is the same for all hadron species. Thus, the extracted $T$ and $\gamma_S$ do not change.
On the other hand, due to the suppression of the densities the total freeze-out volume will be larger compared to the ideal gas.
It is also likely that eigenvolume corrections will not cancel out exactly within the CE formulation.
Still, one expects the system volume to be affected most strongly.
Thus, we fix  $T$ and $\gamma_S$ to the values which were previously
obtained within the ideal HRG model and only vary the system radius $R$.
Three values of the hadron hard-core radius, $r = 0$, 0.3, and 0.5~fm are considered in the MC calculations.
The presence of the strangeness undersaturation parameter $\gamma_S$ is implemented
by the substitution $z_i \to \gamma_S^{|s_i|} \, z_i$ in Eqs.~\eqref{eq:DiagonalMC} and \eqref{ct-F},
where $|s_i|$ is the sum of strange quarks and antiquarks in hadron species $i$.
Note that direct analytic calculation of the average hadron yields from the partition
function is infeasible here due to a very large number of components in the full HRG.
This is quite different from simple systems considered in previous subsections.

The mean multiplicity $\langle N_i \rangle$ is calculated
as a sum of the primordial mean multiplicity
$\langle N^{\rm prim}_i\rangle $ and resonance decay
contributions as follows
\eq{\label{eq:NtotMC}
\langle N_i \rangle
~ =~
\langle N^{\rm prim}_i\rangle~ +~ \sum_R \langle n_i \rangle_R \, \langle N^{\rm prim}_R\rangle~.
}
In contrast to analytic calculations, here the $\langle N^{\rm prim}_i\rangle$ and $\langle N^{\rm prim}_R\rangle$
are calculated by averaging over the sufficiently large number of the weighted events in the MC approach.

The quality of the data description is quantified by $\chi^2$, defined as
\eq{\label{eq:chi2}
\chi^2 = \sum_i \frac{(\langle N_i \rangle - N_i^{\rm exp})^2}{(\sigma_i^{\rm exp})^2},
}
where $i=\pi^+,\,\pi^-,\,K^+,\,K^-,\,\bar{p}$, the $\langle N_i^{\rm exp}\rangle$
and $\sigma_i^{\rm exp}$ are, respectively,
the corresponding experimental yields
and uncertainties, and $\langle N_i\rangle$ is the total yield of hadron species $i$ in the HRG model
calculated with Eq.~(\ref{eq:NtotMC}).

\begin{figure}[t]
\centering
\includegraphics[width=\textwidth]{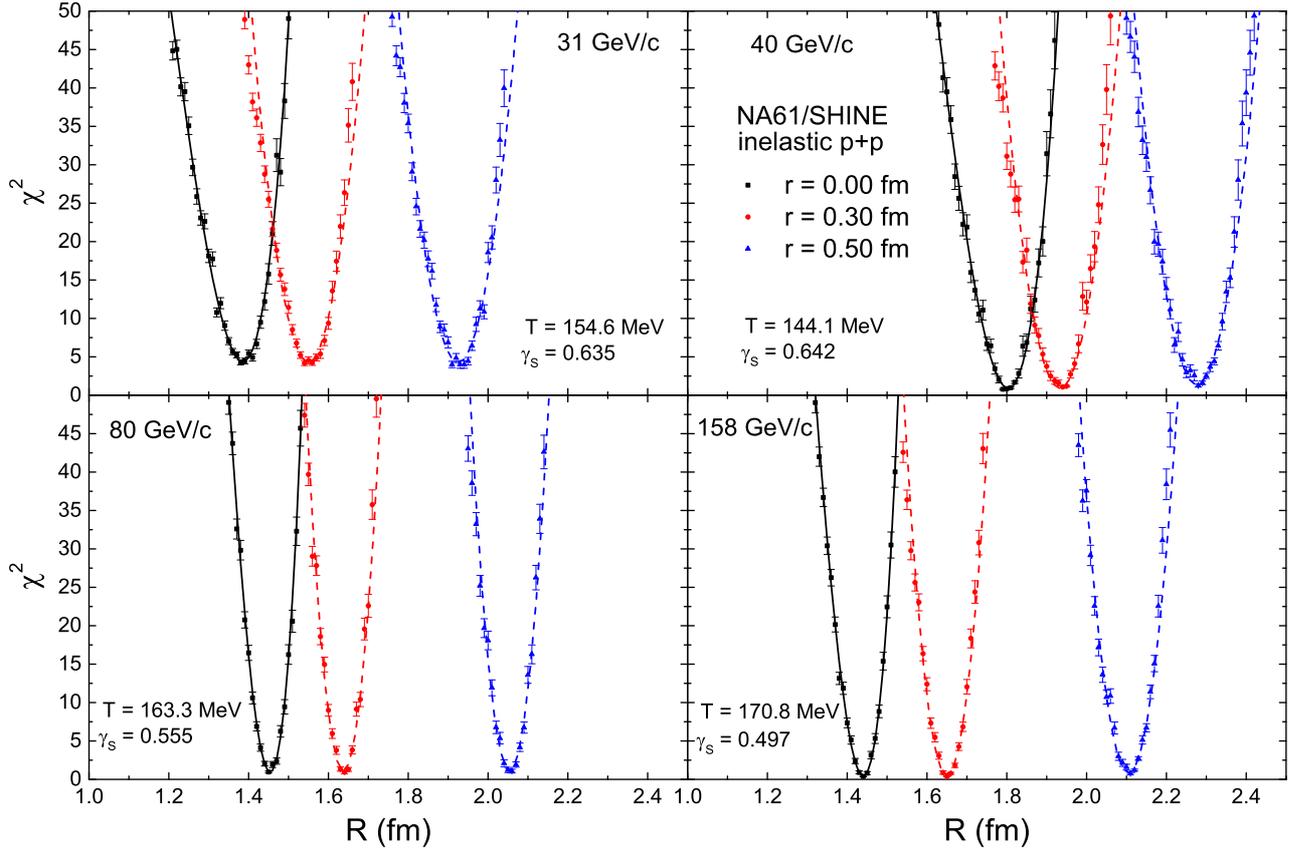}
\caption[]{
The dependence of the $\chi^2$ \eqref{eq:chi2} of the HRG description
of proton-proton hadron yield data of the NA61/SHINE Collaboration
at $p_{\rm lab} = 31,\,40,\,80,$ and $158$~GeV/$c$~\cite{NA61-p+p-mult} on the system radius $R$.
The MC formulation of the eigenvolume HRG in CE is used.
All hadrons are assumed to have the same hard-core radius of
$r = 0$~(black symbols), 0.3~fm (red symbols), and 0.5~fm (blue symbols).
The solid black lines show the results of the analytical calculation
of the $\chi^2$ within the ideal HRG.
The dashed lines depict the parabolic fits to the corresponding MC results in the vicinity of the $\chi^2$ minimum~(see text).
The parameters $T$ and $\gamma_S$ are fixed at each collision energy and are taken from
the ideal HRG model fits performed in Ref.~\cite{VBG2015}.
}\label{fig:NA61fits}
\end{figure}

The MC results for the dependence of the $\chi^2$ on the total
system radius (volume) $R$ are presented in Fig.~\ref{fig:NA61fits}.
The results were obtained by generating $10^5$ weighted events for each
configuration at each considered value of the system radius $R$.
First we note that the MC results for the ideal HRG ($r=0$) are fully consistent with the
corresponding analytic calculations depicted in Fig.~\ref{fig:NA61fits} by solid black lines.
The resulting values of the $\chi^2$ at the global minimum for ideal HRG case are close to those found in Ref.~\cite{VBG2015}.
The MC results for the EV HRG model with $r=0.3$~fm and 0.5~fm are depicted by red and blue symbols, respectively.
We fit our MC results for the $R$-dependence of the $\chi^2$ in the vicinity of the global minimum~(defined as the region where $\chi^2 < 30$) by a parabolic function. 
This allows us to estimate the value and position of the minimum. 
The result of the fits is depicted by dashed lines in Fig.~\ref{fig:NA61fits}.

The minimum values of $\chi^2$ for $r=0.3$~fm and $r=0.5$~fm are very similar to the ones at $r=0$, i.e.
no significant improvement or worsening of thermal fits is observed.
The minima, however, are located at notably higher values of $R$ compared to the ideal HRG model.
This looks very similar to GCE results
where the EV corrections are canceled out in the ratios of yields.
Note, however, that both the temperature $T$ and the $\gamma_S$ parameter were fixed and had the same values at all $R$.
Thus, the $R$-dependencies of the $\chi^2$ shown in Fig.~\ref{fig:NA61fits} should not
be mistaken for the $\chi^2$ profiles of parameter $R$, as neither $T$ nor $\gamma_S$ were fitted at each value of $R$.
One should simultaneously fit all three parameters ($T$, $\gamma_S$, $R$) in order to make a stronger conclusion.
Evidently, the $\chi^2$ profiles may show wider minima.
A more complicated picture can also be expected in EV models with different eigenvolume parameters for different hadron species.
These extensions of the MC calculations
are beyond the scope of the present paper.

\section{Summary}\label{sec-sum}
In summary, we have presented the Monte Carlo (MC) procedure for sampling the hadron yields within the hadron
resonance gas (HRG) model in the grand canonical (GCE) and canonical ensembles (CE).
Both the excluded-volume effects and the exact charge conservation
effects are taken into account simultaneously, with the help of the importance sampling technique.
The MC procedure allows one to calculate arbitrary moments of the event-by-event hadron yields.
To the best of our knowledge, the CE formulation for the excluded-volume HRG had previously been missing.

The MC simulations coincide
with the previously known analytical results for the limiting cases
of the excluded-volume HRG in the GCE for large enough system volumes
and for the CE of non-interacting particles. The MC results for the CE excluded-volume model
are new as these systems were not discussed previously in the literature.
Besides, the finite-size effects
are observed. These effects, usually neglected in the analytical models, exist for both, the particle number densities and the event-by-event fluctuations, in the CE and the GCE.

We have applied the MC procedure within the full HRG model to study
the simultaneous excluded-volume and CE effects on the description of
hadron yields and event-by-event fluctuations.
In particular, it is shown that the excluded-volume and the exact charge conservation effects on the fluctuations of number of positively or negatively charged particles are significant and of similar magnitude. 
Also, the effects of excluded-volume on the CE thermal fits to hadron yield data of the NA61/SHINE Collaboration in proton-proton collisions have been illustrated.

The simultaneous consideration of the effects related to the hadronic interactions and the exact charge conservation is important for analysis of the event-by-event fluctuations measured in heavy-ion collisions.
These effects have to be taken into account for correct interpretation of the data. 
In particular, this concerns 
already the second order (and higher) susceptibilities of net-charge fluctuations,
as well as the higher order susceptibilities
of net-proton fluctuations.
These fluctuation measures
are being measured by the STAR Collaboration~\cite{Adamczyk:2013dal,Adamczyk:2014fia,Luo:2015ewa} in the search for the QCD critical point.

The implementation of the Monte Carlo approach presented here is available within the open source \texttt{Thermal-FIST} package~\cite{ThermalFIST}.

\vspace{0.5cm}

\section*{Acknowledgements}
We are grateful
to Volker Koch and Anton Motornenko
for fruitful discussions.
V.V. acknowledges the support from HGS-HIRe for FAIR.
The work of M.I.G. is supported by the Program of Fundamental Research of the Department of Physics
and Astronomy of the National Academy of Sciences of Ukraine.
H.St. appreciates the support from the J.M.~Eisenberg Laureatus chair.

\begin{appendix}

\section*{Appendix}
%\appendix

This appendix extends the Monte Carlo procedure to the case of the full van der Waals (vdW) equation, i.e. with the presence of both the attractive and repulsive interactions between hadrons. Such an extension permits one to study the important effects related to the nuclear liquid-gas criticality, in particular regarding the higher moments of the conserved charges fluctuations~\cite{Fukushima:2014lfa,Vovchenko:2016rkn}.

The pressure of a multi-component Boltzmann system with the vdW interactions reads~\cite{Vovchenko:2017zpj}
\eq{\label{eq:pvdw}
p(T,n_1,\ldots,n_f)~ =~ \sum_{i=1}^f \frac{T\,n_i}{1-\sum_j \tilde{b}_{ji}\,n_j} ~-~ \sum_{i,j=1}^f a_{ij} \, n_i \, n_j~.
}
Here the parameters $\tilde{b}_{ji}$ correspond to the repulsive vdW interactions and have the same physical meaning as in the NDE model in Sec.~\ref{subsec-NDE}.
The parameters $a_{ij}$ correspond to the attractive vdW interactions, for each pair of particle species.

The pressure \eqref{eq:pvdw} corresponds to the following GCE partition function
\eq{\label{Zvdw}
\mathcal{Z}_{\rm vdW}(V,T,\mu_1,\ldots,\mu_f) & =
\sum_{N_1=0}^{\infty} \ldots \sum_{N_f=0}^{\infty} \, 
 \prod_{i=1}^f \,
\exp\left(\frac{\mu_i N_i}{T}\right)~
\frac{[(V-\sum_{j=1}^f \tilde{b}_{ji} N_j)\,z_i]^{N_i}}{N_i!} \, \exp\left(\sum_{j=1}^f \, \frac{a_{ij} N_j}{VT} N_i\right)
\nonumber \\
& \quad \times \theta(V-\sum_{j=1}^f \tilde{b}_{ji} N_j)~,
}

The CE partition function is obtained by introducing the corresponding Kronecker delta functions~(see Sec.~\ref{sec-CE} for details):
\eq{
\label{eq:vdWCE}
 Z_{\rm vdW}(V,T,\{Q\})  ~ &=~ \sum_{N_1=0}^{\infty} \ldots \sum_{N_f=0}^{\infty} \, 
 \prod_{i=1}^f \,
 \frac{\left[ (V - \sum_{j=1}^f \tilde{b}_{ji} N_j) \,  z_i
\right]^{N_i}}{N_i!}\, 
\exp\left(\sum_{j=1}^f \, \frac{a_{ij} N_j}{VT} N_i\right)  \,
\nonumber \\
&  \quad \times ~\theta(V - \sum_{j=1}^f \tilde{b}_{ji} N_j)\, \prod_{k=1}^c \delta(Q_k - \sum_j q_k^{(j)} N_j)~.
}

The $F$ and $\Theta$ functions, which define the MC procedure described in Sec.~\ref{sec-MC}, are the following:
\eq{
F_{\rm vdW}(N_1, \ldots, N_f;V,T,\{\mu_Q\}) & ~=~ \prod_{i=1}^{f} \, \frac{\left[ (V - \sum_j \tilde{b}_{ji} N_j) \,  z_i
\, e^{\mu_i/T} \, e^{\sum_{j=1}^f \, \frac{a_{ij} N_j}{VT}}
\right]^{N_i}}{N_i!}~, \label{ct-F}\\
\Theta_{\rm vdW}(N_1, \ldots, N_f;V) & ~=~ \prod_{i=1}^{f} \, \theta(V - \sum_j \tilde{b}_{ji} N_j)~.
}

\end{appendix}

\end{document}